\documentclass[12pt,twoside,a4paper]{article}
\usepackage{amsmath,amssymb,latexsym,theorem,natbib,epsfig,color,subfigure, dsfont}
\usepackage{multirow,graphicx,array,rotating,appendix}  
\usepackage{epstopdf,url,afterpage}
\usepackage[breaklinks, colorlinks=true, citecolor=black, urlcolor=black, linkcolor=black]{hyperref}

\newfont{\msa}{msam10 scaled\magstep1}
\newfont{\ssmsa}{msam9}

\def\crps{\mathop{\hbox{\rm CRPS}}}

\def\bs{\mathop{\hbox{\rm BS}}}

\def\clzero{{CL0}}

\numberwithin{equation}{section}

\title{Post-processing numerical weather prediction ensembles for probabilistic solar irradiance forecasting}

\author{ {\sc Benedikt Schulz$^1$}, {\sc Mehrez El Ayari$^{2,3}$}, {\sc Sebastian Lerch$^{1,4}$}\\ and {\sc S\'andor Baran$^2$} \vspace*{0.5cm}\\
         $^1$Institute for Stochastics, Karlsruhe Institute of Technology, \\
         Englerstra\ss{}e 2, D-76131 Karlsruhe, Germany \\
         $^2$Faculty of Informatics, University of Debrecen\\
         Kassai \'ut 26, H-4028 Debrecen, Hungary\\
         $^3$Doctoral School of Informatics, University of Debrecen\\
         Kassai \'ut 26, H-4028 Debrecen, Hungary \\
         $^4$Heidelberg Institute for Theoretical Studies, \\
         Schloss-Wolfsbrunnenweg 35, D-69118 Heidelberg, Germany
        }

\date{}
\begin{document}
\pagestyle{myheadings}

\maketitle

\begin{abstract}
In order to enable the transition towards renewable energy sources, probabilistic energy forecasting is of critical importance for incorporating volatile power sources such as solar energy into the electrical grid. Solar energy forecasting methods often aim to provide probabilistic predictions of solar irradiance. In particular, many hybrid approaches combine physical information from numerical weather prediction models with statistical methods. Even though the physical models can provide useful information at intra-day and day-ahead forecast horizons, ensemble weather forecasts from multiple model runs are often not calibrated and show systematic biases. We propose a post-processing model for ensemble weather predictions of solar irradiance at temporal resolutions between 30 minutes and 6 hours. The proposed models provide probabilistic forecasts in the form of a censored logistic probability distribution for lead times up to 5 days and are evaluated in two case studies covering distinct physical models, geographical regions, temporal resolutions, and types of solar irradiance. We find that post-processing consistently and significantly improves the forecast performance of the ensemble predictions for lead times up to at least 48 hours and is well able to correct the systematic lack of calibration.

\end{abstract}

\section{Introduction}
\label{sec1}

To reduce emissions of greenhouse gases a transition towards renewable energy sources such as wind and solar power is imperative \citep{vanderMeerEtAl2018}. Accurate and reliable forecasts of power generation from those sources are thus becoming increasingly important for integrating volatile power systems into the electrical grid in order to balance demand and supply \citep{GottwaltEtAl2016,GonzalesOrdianoEtAl2020}.
The literature on energy forecasting has primarily focused on deterministic prediction for the past decades. However, it has now been widely argued that probabilistic forecasting is essential for optimal decision making in planning and operation \citep{HongFan2016,vanderMeerEtAl2018,HauptEtAl2019,HongEtAl2020}. For example, \citet{HongEtAl2020} identify probabilistic forecasting with the aim of providing a predictive probability distribution for a future quantity or event in order to quantify forecast uncertainty as one of the most important emerging research topics in their recent review on energy forecasting.

We here focus on solar energy which is one of the most important sources of renewable energy. For example, photovoltaic (PV) power contributes significantly to the power supply in Germany and generated  8.2\%  of the  gross  electricity  consumption in 2019, and temporarily up to 50\% of the current electricity consumption on sunny days \citep{FraunhoferISE}.

Solar energy forecasting approaches can be distinguished into those that aim to predict solar irradiance, and those that aim to predict PV power. Naturally, solar irradiance and PV system output are strongly correlated, and the employed statistical methods are similar \citep{vanderMeerEtAl2018}. We will focus on probabilistic solar irradiance forecasting in the following. 

For recent comprehensive overviews and reviews of existing approaches, see \citet{vanderMeerEtAl2018} and \citet{Yang2019ROPES}. Except for short-term prediction \citep[e.g.,][]{ZelikmanEtAl2020}, most methods for probabilistic solar irradiance forecasting combine physical information from numerical weather prediction (NWP) models with statistical methods. NWP models describe atmospheric processes via systems of partial differential equations and are often run several times with varying initial conditions and model physics, resulting in an ensemble of predictions that provide a probabilistic forecast \citep{BauerEtAl2015}. Despite significant advances over the past decades ensemble forecasts continue to exhibit systematic errors that require correction. To that end, a rapidly growing collection of methods for statistical post-processing has been developed in the meteorological and statistical literature. Post-processing methods can be viewed as distributional regression models that produce probabilistic forecasts of a target variable (such as solar irradiance), using ensemble predictions of (potentially many) meteorological variables as input. 
Prominent methods for post-processing include the ensemble model output statistics (EMOS) approach proposed in \citet{GneitingEtAl2005}, where the forecast distribution is given by a parametric probability distribution with parameters connected to the ensemble predictions of the variable of interest via link functions. Other widely used post-processing techniques build on quantile regression methods which provide nonparametric probabilistic forecasts in the form of quantiles of the forecast distribution. 
We refer to \citet{wilks18} for a general introduction and to \citet{VannitsemEtAl2020} for a comprehensive overview of current research developments and operational implementations. 

Post-processing ensemble predictions of solar irradiance has recently received a growing interest in the literature on probabilistic solar energy forecasting\footnote{Note that related post-processing methods have also been applied for direct forecasting of PV power output, for example in \citet{SperatiEtAl2016}. We limit our literature review to methods that aim to predict solar irradiance to retain focus.}. \citet{BakkerEtAl2019} compare post-processing methods for the clear-sky index that take deterministic NWP predictions of several variables as input, and employ various machine learning approaches for quantile regression. \citet{LaSalleEtAl2020} propose an EMOS model for global horizontal irradiance where truncated normal and generalized extreme value distributions are used as forecast distributions, and compare to quantile regression and analog ensemble methods. \citet{YagliEtAl2020} compare several parametric and nonparametric post-processing methods for hourly clear-sky index forecasting, including EMOS models based on Gaussian, truncated logistic and skewed Student's t distributions as well as quantile regression based on random forests, and generalized additive models for location, scale and shape. In closely related work, \citet{Yang2020siteadaptation1} proposes the use of EMOS models for probabilistic site adaptation of gridded solar irradiance products, and \citet{Yang2020siteadaptation2} compares building models for irradiance and for the clear-sky index and investigates the choice of parametric distributions. A comprehensive overview of post-processing methods in solar forecasting is provided in the recent review paper of \citet{YangVanDerMeer2021}.

Here, we build on the EMOS framework and propose post-processing models for different measured types of solar irradiance, using corresponding NWP ensemble predictions as input.
Within the classification introduced in \citet{YangVanDerMeer2021}, this corresponds to a \textit{probabilistic-to-probabilistic} post-processing method.
To account for the specific discrete-continuous nature of solar irradiance due to the positive probability of observing zero irradiance during night-time, we use a censored logistic forecast distribution motivated by similar models for post-processing ensemble forecasts of precipitation accumulation \citep{sch14,bn16}. The post-processing models are applied in two case studies that focus on distinct solar irradiance variables, NWP models, temporal resolutions and geographic regions (Hungary and Germany), for lead times of up to 48 and 120 hours, respectively. We utilize periodic models to better capture seasonal variation in solar irradiance, and investigate different temporal compositions of training datasets for model estimation. Further, we compare the effects using a clear-sky index as target variable to stationarize the time series of irradiances -- the standard approach in solar forecasting -- and post-processing the irradiance forecasts directly.  

The remainder of the paper is organized as follows. Section \ref{sec2} introduces the datasets used in the case studies. The proposed post-processing approach, training procedures and forecast evaluation methods are described in Section \ref{sec3}, with results for the two case studies presented in Section \ref{sec4}. The paper concludes with a discussion in Section \ref{sec5}, additional results are  deferred to the Appendix.

\section{Data}
\label{sec2}

In the case studies of Section \ref{sec4}, we evaluate the EMOS models proposed in Section \ref{subs3.2} using forecasts of different types of solar irradiance produced by two different ensemble prediction systems (EPSs), which cover distinct forecast domains.

\subsection{AROME-EPS}
\label{subs2.2}
The 11-member Applications of Research to Operations at Mesoscale EPS (AROME-EPS) of the Hungarian Meteorological Service (HMS) covers the Transcarpatian Basin with a horizontal resolution of 2.5 km \citep{rvsz20}. It consists of 10 ensemble members obtained from perturbed initial conditions and a control member from an unperturbed analysis. The dataset at hand contains ensemble forecasts of instantaneous values of global horizontal irradiance (GHI) ($W/m^2$) together with the corresponding validation observations of the HMS for seven representative locations in Hungary (Asz\'od, Budapest, Debrecen, Kecskem\'et, P\'ecs, Szeged, T\'api\'oszele) for the period between 7 May 2020 and 14 October 2020.  Forecasts are initialized at 00 UTC with a forecast horizon of 48h and a temporal resolution of 30 minutes resulting in a total of 96 forecasts per submission. 
Rephrasing this in the terminology of \citet{Yang2020time}, the temporal information can be written as
$$\lbrace \mathcal{S}^{48\text{h}}, \mathcal{R}_f^{30\text{min}}, \mathcal{L}^{0\text{h}}, \mathcal{U}^{24\text{h}} \rbrace \quad \text{and} \quad \mathcal{R}_f = \mathcal{R}_x, $$
where $\mathcal{S}$ refers to the forecast span covered by all forecasts, $\mathcal{R}_f$ to the temporal resolution of the forecasts, $\mathcal{R}_x$ to that of the observations, $\mathcal{L}$ to the time between submission and first time stamp, and $\mathcal{U}$ to the time interval between consecutive submissions.

For the AROME-EPS, we will refer to the term lead time as the time between the initialization and the time stamp of the corresponding instantaneous forecast.

\subsection{ICON-EPS}
\label{subs2.1}

The 40-member global ICOsahedralNonhydrostatic EPS \citep[ICON-EPS;][]{zrpb15} of the German Meteorological Service (DWD; Deutsche Wetterdienst) was launched in 2018 and has a horizontal resolution of 20 km over Europe (ICON-EU EPS). 
The ensemble members are generated with the help of random perturbations and the forecasts are initialized four times a day at 00/06/12/18 UTC each with a forecast horizon of 120h \citep{icon}. 
Within the first 48h the temporal resolution is 1h, 3-hourly up to 72h, while between 72 and 120h the resolution is 6h (a total of 64 forecasts). Adapting the terminology of \citet{Yang2020time} accordingly, this corresponds to
$$\lbrace \mathcal{S}^{(48\text{h}, 24\text{h}, 48\text{h})}, \mathcal{R}_f^{(1\text{h}, 3\text{h}, 6\text{h})}, \mathcal{L}^{(0\text{h}, 48\text{h}, 72\text{h})}, \mathcal{U}^{6\text{h}} \rbrace. $$
The ICON ensemble predictions are given as averages between two time stamps, e.g.\ the 12-step ahead forecast is the average predicted irrandiance between 11 to 12h after initialization time and the 59-step ahead forecast is the average from 84 to 90h. For simplicity, we will refer to a individual forecast by the lead time and not the step ahead, where lead time refers to the time between submission and the final time stamp, i.e., the former forecast has a lead time of 12h, the latter a lead time of 90h.

Our dataset contains ensemble forecasts of the two components of GHI: beam normal irradiance (BNI) adjusted for the solar zenith angle $\theta$ (i.e., $\text{BNI}\cdot\cos({\theta})$), and diffuse horizontal radiation (DHI) ($W/m^2$). 
To simplify the distinction between the different types of irradiance and improve the readability of the article, we will refer to $\text{BNI}\cdot\cos({\theta})$ as beam horizontal irradiance (BHI) or direct irradiance, to DHI as diffuse irradiance, and to $\text{GHI} = \text{BHI} + \text{DHI}$ as global irradiance, in the following.

We further obtained corresponding observational data for weather stations located near the major cities of Berlin, Hamburg and Karlsruhe from Open Data Server of DWD \citep{DWDCDC}.
The observations are computed based on 10-minute sums of the corresponding variables, i.e.\ the temporal resolution of the observation is $\mathcal{R}_x^{10\text{min}}$. For detailed descriptions, we refer to \citet{icon} for the ensemble predictions and \citet{BeckerBehrens2012} for the observations.
The entire dataset used here covers the period 27 December 2018 -- 31 December 2020.

\section{Post-processing methods and forecast evaluation}
\label{sec3}

Broadly, two kinds of statistical post-processing methods can be distinguished. Parametric (or distribution-based) methods specify the distribution of the variable of interest, whereas nonparametric (or distribution-free) methods do not assume a certain forecast distribution and typically forecast a set of quantiles or random draws that represent the distribution. 
While distribution-free methods do not require the choice of a parametric family of probability distributions for the forecast model, they usually do not offer a complete representation of the forecast distribution. For example, quantile regression approaches only provide predictions for a set of quantiles, and it might not be clear how to extrapolate beyond the range of the considered quantile levels. By contrast, parametric methods provide full predictive distributions from which arbitrary quantities, e.g.\ quantiles, can be computed. In addition, parametric methods are generally better suited for situations in which only limited amounts of training data are available. In view of the sparsity of data, we here choose the distribution-based EMOS approach. In the following, we propose a novel EMOS model based on a censored logistic distribution.

\subsection{Choice of forecast distribution}
  \label{subs3.1}

The discrete-continuous nature of solar irradiance calls for non-negative predictive distributions assigning positive mass to the event of zero irradiance. Similar to parametric approaches to post-processing ensemble forecasts of precipitation accumulation, one can either left-censor an appropriate continuous distribution at zero \citep[see e.g.][]{sch14,bn16}, or choose the more complex method of mixing a point mass at zero and a suitable continuous distribution with non-negative support \citep{srgf07,bf12}. Here, we focus on the former and introduce the censored logistic distribution on which our approach is built.

Consider a logistic distribution \ ${\mathcal L}(\mu,\sigma)$ \ with location \ $\mu$ \ and  scale \ $\sigma>0$ \ specified by the probability density function (PDF)
\begin{equation*}
  g(x;\mu,\sigma ):=\frac {{\mathrm e}^{-(x-\mu)/\sigma}}{\sigma\left(1+{\mathrm e}^{-(x-\mu)/\sigma}\right)^2},  \qquad x\in{\mathbb R},
\end{equation*}  
and the cumulative distribution function (CDF) \ $G(x;\mu,\sigma):=\big(1+{\mathrm e}^{-(x-\mu)/\sigma}\big)^{-1}$.

The logistic distribution left-censored at zero (\clzero) assigns point mass \ $G(0;\mu,\sigma)=\big(1+{\mathrm e}^{\,\mu/\sigma}\big)^{-1}$ \ to the origin, i.e.\ the probability of observing a negative value (before censoring) is the probability of observing zero afterwards.
The \clzero-distribution can be defined by the CDF
\begin{equation}
  \label{eq:cl0_cdf}
  G_0^c(x;\mu,\sigma):=\begin{cases} G(x;\mu,\sigma),& \quad x\geq 0, \\
    0, & \quad x<0. \end{cases}
\end{equation}
or the generalized PDF
\begin{equation}
  \label{eq:cl0_pdf}
  g_0^c(x;\mu,\sigma )={\mathds 1}_{\{x=0\}}G(0;\mu,\sigma) + {\mathds 1}_{\{x>0\}}g(x;\mu,\sigma ),
\end{equation}
where \ $\mathds 1_A$ \ denotes the indicator function of a set \ $A$. \ The \ $p$-quantile \ $q_p$ \ $(0<p<1)$ \ of \eqref{eq:cl0_cdf} equals \ $0$ \ if \ $p\leq G(0;\mu,\sigma)$ \ and \ $q_p=\mu-\sigma \big(\log(1-p)-\log p\big)$, \ otherwise. With the help of \eqref{eq:cl0_pdf} it is straightforward to show that the corresponding mean equals
\begin{equation*}
  \mu_0^c:=\mu+\sigma \log \big(1+{\mathrm e}^{-\mu/\sigma}\big).
\end{equation*}

\begin{figure}
  \centering
\epsfig{file=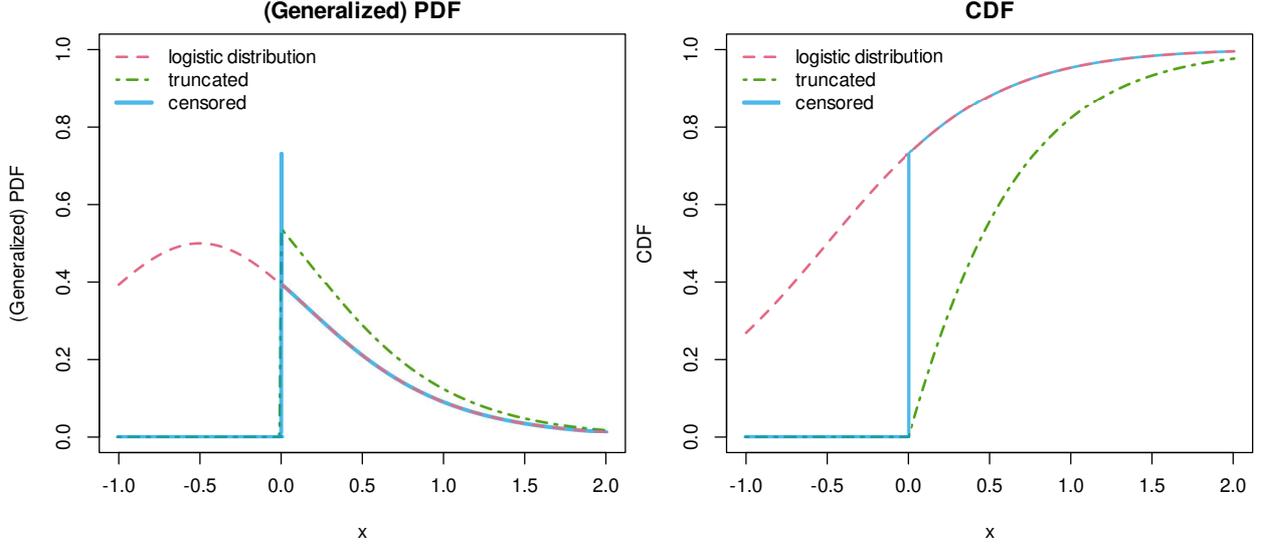, width=\textwidth}
\caption{Illustration of the (generalized) PDF and CDF of a logistic distribution with parameters $\mu = -0.5$ and $\sigma = 0.5$, and the effects of truncating and left-censoring at 0.}
\label{fig:distr_illustration}
\end{figure}

Note that initial tests with a censored normal predictive distribution were performed for the ICON-EPS dataset; however, the results suggested that the proposed \clzero-EMOS approach results in slightly improved predictive performance. The choice of parametric families for the forecast distribution has been an important aspect in post-processing research. For considerations in the context of solar irradiance forecasting, see e.g.\  \citet{Yang2020siteadaptation2}, \citet{YagliEtAl2020}, and \citet{LaSalleEtAl2020}. 
In those previous works, forecast distributions are truncated at zero, i.e., probability mass belonging to negative values before truncation is re-assigned to positive values by restricting the support to the positive half-axis and multiplying the PDF with a constant factor. The key difference of left-censoring at zero is given by the positive probability assigned to the event of observing exactly zero irradiance. Figure \ref{fig:distr_illustration} illustrates this difference for a truncated and left-censored logistic distribution.

Our choice of a left-censored distribution was motivated by the aim to obtain a single distributional model that can be applied to all times of day and is able to account for cases where the observation and all or most of the ensemble member predictions are zero, which makes it unnecessary to remove night-time irradiance data during training and inference. Therefore, there is no need to select location- and season-specific times of day that define periods of time where the post-processing model can be applied, which is the case for models based on truncated distributions. In addition, we have observed occasional cases where zero irradiance is observed, but some of the ensemble members predict non-zero values. A model based on a censored distribution is able to correct those deficiencies and might have advantages for applications in automated procedures where post-processing constitutes one of the components and post-processed forecasts serve as inputs for additional modeling steps.

Generally, it has been noted that a single parametric distribution may not result in a perfect fit, and more involved approaches based on mixtures or combinations of several forecast distributions that have been proposed in the meteorological literature might be able to improve performance, but increase model complexity and computational costs \citep{BaranLerch2016,BaranLerch2018}.

\subsection{Ensemble model output statistics models for solar irradiance forecasting}
  \label{subs3.2}

Now, denote by \ $f_1,f_2, \ldots ,f_K$ \ the ensemble member forecasts of solar irradiance for a given location, time point and forecast horizon. In the simplest proposed EMOS model, the location parameter \ $\mu$ \ and the scale parameter \ $\sigma$ \ of the \clzero-distribution are connected to the ensemble members via link functions
\begin{equation}
  \label{eq:cl0link1}
  \mu = \alpha_0+\alpha_1f_1+ \cdots + \alpha_Kf_K + \nu p_0 \qquad \text{and} \qquad \sigma = \exp\big(\beta_0 + \beta_1 \log S^2\big),
\end{equation}
where \ $p_0$ \ and \ $S^2$ \ are the proportion of zero observations and the ensemble variance, respectively, that is
\begin{equation*}
  p_0:=\frac 1K\sum_{k=1}^K{\mathds 1}_{\{f_k=0\}} \qquad \text{and} \qquad S^2:=\frac 1{K-1}\sum_{k=1}^K(f_k-\overline f)^2,
  \end{equation*}
with \ $\overline f$ \ denoting the ensemble mean. The EMOS coefficients \ $\alpha_0,\alpha_1, \ldots ,\alpha_K, \nu$ \ and \ $\beta_0,\beta_1$ \ are estimated according to the optimum score principle of \citet{grjasa07}, i.e., by optimizing the mean value of an appropriate verification metric over the training data consisting of past pairs of forecasts and observations for a given time period.

In order to capture the seasonal variation in solar irradiance, following the ideas of \citet{hemri14}, we further fit separate periodic models to both observations and ensemble forecasts of the training data. Two regression models dealing with oscillations of a single and two different frequencies are investigated, namely
  \begin{align}
    \label{eq:reg1}
    y_t&=c_0+c_1\sin \Big(\frac{2\pi t}{365}\Big) + c_2\cos \Big(\frac{2\pi t}{365}\Big)+\varepsilon _t \qquad \qquad \text{and} \\
    y_t&=d_0+d_1\sin \Big(\frac{2\pi t}{365}\Big) + d_2\cos \Big(\frac{2\pi t}{365}\Big)+d_3\sin \Big(\frac{4\pi t}{365}\Big) + d_4\cos \Big(\frac{4\pi t}{365}\Big)+\varepsilon _t, \label{eq:reg2}
    \end{align}
where the dependent variables \ $y_t, \ t=1,2, \ldots ,n,$ \ are either irradiance observations for a given location or members of the corresponding ensemble forecast with a given lead time \ $h$ \ from a training period of length \ $n$. \ With the help of either \eqref{eq:reg1} or \eqref{eq:reg2} one can calculate the \ $h$ \ ahead predictions \ $\widehat y$ \ and \ $\widehat f_k$ \ of the observation and ensemble members, respectively, and consider the following modified link function for the location:
\begin{equation}
  \label{eq:cl0link2}
  \mu = \widehat y + \alpha_0+\alpha_1\big(f_1-\widehat f_1\big)+ \cdots + \alpha_K\big(f_K-\widehat f_K\big) + \nu p_0.
\end{equation}

Model formulations \eqref{eq:cl0link1} and \eqref{eq:cl0link2} are valid under the assumption that each ensemble member can be identified and tracked. However, most operationally used EPSs today generate ensemble forecasts that lack individually distinguishable physical features such as distinct variations in the model physics, for example by generating ensemble member based on random perturbations of initial conditions. Those statistically indistinguishable members (or groups of members) generated in this way are usually referred to as exchangeable \citep{FraleyEtAL2010} in reference to the concept of exchangeable random variables in statistics. This is also the case for the ICON-EPS and the AROME-EPS described in Section \ref{sec2}. The existence of groups of exchangeable ensemble members should be taken into account during model formulation. This is usually achieved by requiring that ensemble members within a given group share the same coefficients \citep[see e.g.][]{wilks18}. If there exist \ $M$ \ ensemble members divided into \ $K$ \ exchangeable groups and \ $\overline f_k$ \ denotes the mean of the $k$th group containing \ $M_k$ \ ensemble members \ ($\sum_{k=1}^KM_k=M$), the exchangeable versions of link functions \eqref{eq:cl0link1} and \eqref{eq:cl0link2} are
\begin{equation}
  \label{eq:cl0link1Ex}
  \mu = \alpha_0+\alpha_1\overline f_1+ \cdots + \alpha_K\overline f_K + \nu p_0
\end{equation}
and
\begin{equation}
  \label{eq:cl0link2Ex}
  \mu = \widehat y + \alpha_0+\alpha_1\big(\overline f_1-\widetilde f_1\big)+ \cdots + \alpha_K\big(\overline f_K-\widetilde f_K\big) + \nu p_0,
\end{equation}
respectively, where \ $\widetilde f_k$ \ is the prediction of \ $\overline f_k$ \ for lead time \ $h$ \  based either on \eqref{eq:reg1} or \eqref{eq:reg2}.
      
\subsection{Training data selection}
\label{subs3.3}

The parameters of the \clzero-EMOS model are estimated with the help of ensemble forecasts and corresponding observations from a training data set, where several options in terms of both spatial and temporal composition can be considered. From the spatial point of view, there are two traditional approaches: local and regional (sometimes also called global) selection \citep{tg10}. In the local approach, the parameters of the predictive distribution for a given location are estimated using only data from that particular location, resulting in different parameter estimates for the different locations. In order to ensure numerical stability of the estimation process, local modeling requires long time periods for training, which is the major disadvantage of this approach. As it addresses the location-specific forecast error characteristics, it often results in better forecast skill than regional estimation, where training data of the whole ensemble domain are used and all locations share the same set of parameters.

Regarding the temporal composition, the standard approach in EMOS modeling is the use of rolling training periods, where training data consists of forecasts and observations for the \ $n$ \ calendar days preceding the target date of interest. Rolling training periods can be flexibly applied to smaller datasets and enable models to adapt to changes in meteorological conditions or the underlying NWP system. An alternative approach is to utilize all available data by considering expanding training periods, motivated by studies suggesting that using long archives of training data irrespective of potential NWP model changes during that period often show superior performance \citep{LangEtAl2020}. Regularly extending training sets may for example be relevant in operational implementations where data archives might be built up and expanded over time. In the case studies of Section \ref{sec4}, examples of all listed training data selection methods are shown: regional estimation with a rolling training period in Section \ref{subs4.1} and local estimation with rolling and extending training periods in Section \ref{subs4.2}. 

For both datasets, ensemble predictions of multiple lead times are available, and in case of the ICON-EPS dataset the forecasts are initialized by the NWP model at four different times of the day. These are treated separately when estimating model parameters, i.e., a separate post-processing model is estimated for each lead time and each initialization hour, based on training datasets comprised of data from those lead times and initialization hours only. Thereby, we aim to account for changes in the forecast error characteristics of the raw ensemble predictions over multiple lead times, and for potential diurnal effects by ensuring that the training data covers the same time of day of the observation. Since we do not remove night-time data during training and inference, this further helps to account for positive probabilities of observing zero irradiance as point masses in the forecast distributions. Note that seasonal variations for a given time of day, for example effects of differing solar zenith angles, are implicitly modeled when using \eqref{eq:reg1} or \eqref{eq:reg2}. 

\subsection{Forecast evaluation}
\label{subs3.4}

As mentioned in Section \ref{subs3.2}, the estimates of the unknown parameters of the \clzero-EMOS model minimize the mean of a proper verification score over the training data. These so-called proper scoring rules \citep{grjasa07} have been widely used in the meteorological and economic literature, and have recently become popular tools for evaluating probabilistic solar energy forecasts. See, for example, the recommendations in \citet{LauretEtAl2019} which we will follow throughout. 

The most popular proper scoring rule in the atmospheric sciences is the continuous ranked probability score \citep[CRPS;][Section 9.5.1]{wilks19}, which simultaneously assesses both calibration and sharpness of the probabilistic forecasts. The former refers to a statistical consistency between forecasts and observations, whereas the latter refers to the concentration of the predictive distribution, see \citet{LauretEtAl2019} for details. For a predictive CDF \ $F$ \ and real-valued observation \ $x$, \ the CRPS is defined as
\begin{equation}
  \label{eq:CRPS}
\crps\big(F,x\big):=\int_{-\infty}^{\infty}\big (F(y)-{\mathds 1}_{\{y \geq x\}}\big )^2{\mathrm d}y={\mathsf E}|X-x|-\frac 12
{\mathsf E}|X-X'|, 
\end{equation}
where \ $X$ \ and \ $X'$ \ are independent random variables with CDF \ $F$ \ and finite first moment. The CRPS is a negatively oriented score, i.e., smaller values indicate better forecasts, and the second representation in \eqref{eq:CRPS} implies that it can be expressed in the same unit as the observation. Further, it can be seen as a generalization of the absolute error to probabilistic forecasts and allows comparisons with point forecasts. For the \clzero-distribution defined by \eqref{eq:cl0_cdf}, the CRPS can be computed in closed form \citep[for the exact formula see][]{jkl19}, which enables computationally efficient parameter estimation.
For a given lead time, the forecast skill of competing probabilistic forecasts is compared based on the mean CRPS over all corresponding forecast cases in the verification data.

For assessing the predictive performance of the different forecasts with respect to the binary event that observation \ $x$ \ exceeds a given threshold \ $z$, \ one can consider the Brier score \citep[BS;][Section 9.4.2]{wilks19}, which for a predictive CDF \ $F$ \ is defined as
\begin{equation}
  \label{eq:BS}
 \bs \big(F,x;z\big):= \big (F(z)-{\mathds 1}_{\{z \geq x\}}\big )^2
\end{equation}
\citep[see e.g.][]{gr11}. To compute the Brier Score, the continuous observation \ $x$ \ is thus converted to a binary threshold exceedance $\ {\mathds 1}_{\{z \geq x\}}\ $ with corresponding predicted probability $\ F(z).\ $ Note that the Brier score is also negatively oriented and the CRPS is the integral of the BS over all possible thresholds $\ z\in\mathbb{R}.\ $ For ensemble forecasts, both scores are defined based on the empirical CDF.

For a given probabilistic forecast \ $F$, \ the improvement in a score \ ${\mathcal S}_F$ \ with respect to a reference forecast \ $F_{\text{ref}}$ \ can be quantified with the help of the corresponding skill score defined as
$$\mathcal{SS}_F:=1-\frac {\overline {\mathcal S}_F} {\overline {\mathcal S}_{F_{\text{ref}}}},$$
where \ $\overline {\mathcal S}_F$ \ and \ $\overline {\mathcal S}_{F_{\text{ref}}}$ \ denote the mean score values over the verification data corresponding to forecasts  \ $F$ \ and  \ $F_{\text{ref}}$, \ respectively. In the case studies, we consider the continuous ranked probability skill score (CRPSS) and the Brier skill score (BSS), and will usually use the raw ensemble forecasts as reference model (except for Section \ref{secB} of the Appendix).

A simple tool of assessing the calibration of ensemble forecasts is the verification rank histogram displaying the histogram of ranks of observations with respect to the corresponding ordered ensemble forecasts \citep[][Section 9.7.1]{wilks19}. For a calibrated $K$-member ensemble, the ranks should be uniformly distributed on the set \ $\{1,2, \ldots ,K+1\}$. \ In the case of continuous forecasts specified by predictive distributions, one can either consider the verification rank histogram of simulated ensembles sampled from the predictive PDF, or investigate the probability integral transform (PIT) histogram \citep[][Section 9.5.4]{wilks19}, where the PIT is the value of the predictive CDF evaluated at the validating observation. For a calibrated predictive distribution, the PIT follows a uniform distribution on the \ $[0,1]$ \ interval. The PIT histogram can thus be considered a continuous counterpart of the verification rank histogram.

Further, one can also investigate the calibration of probabilistic forecasts with the help of the coverage of  \ $(1-\alpha)100\,\%, \ \alpha\in(0,1)$, \ central prediction intervals. Coverage is defined as the proportion of validating observations located between the lower and upper \ $\alpha/2$ \ quantiles of the predictive distribution, which for a calibrated probabilistic forecast should be around \ $(1-\alpha)100\,\%$. \ In order to allow a direct comparison with the raw ensemble, the level \ $\alpha$ \ is usually chosen to match the nominal coverage of the ensemble range, which for a $K$-member ensemble equals \ $(K-1)/(K+1)100\,\%$.

Finally, point forecasts such as median and mean of the raw ensemble and of the predictive distribution are evaluated with the help of the mean absolute error (MAE) and root mean square error (RMSE). Note that the former is optimal for the median, whereas the latter for the mean \citep{gneiting11}.

\section{Case studies}
\label{sec4}

In the following case studies, the forecast skill of various variants of the \clzero-EMOS model introduced in Section \ref{subs3.2} is evaluated. First, we consider a simple \clzero-EMOS variant for the HMS AROME-EPS ensemble forecasts of GHI, then we will investigate the performance of the more complex models for the DWD ICON-EPS ensemble forecasts of direct and diffuse irradiance.

\subsection{Results for the AROME-EPS dataset}
\label{subs4.1}

As discussed in Section \ref{subs2.2}, the AROME-EPS consists of a control member and 10 exchangeable ensemble members obtained using perturbed initial conditions. The dataset at hand covers a short time period only, in particular compared to the ICON-EPS dataset, with forecast-observation pairs available for only 159 calendar days. 
Therefore, the available training periods cannot be long enough for accurate modeling of seasonal oscillations and we only consider a \clzero-EMOS model where the location is linked to the ensemble members via \eqref{eq:cl0link1Ex} with \ $K=2$ \ and \ $M_1=1, \ M_2=10$, \ which means that six  parameters need to be estimated. We consider regional estimation with a rolling training period of length 31 days, leaving 127 calendar days (9 June 2020 -- 13 October 2020) for forecast verification, and refer to this model as the \textit{simple RT} model. The choice of the training period length corresponds to typical values in the post-processing literature and was made to have a similar forecast case per parameter ratio as for the best performing model of Section \ref{subs4.2}. In light of the limited size of the dataset, it is not surprising that the use of monthly expanding training periods or local parameter estimation procedures results in worse predictive performance, and we omit the corresponding results in the interest of brevity. Recall that the ensemble predictions of GHI are provided at a temporal resolution of 30 minutes. As all AROME-EPS forecasts are initialized at 00 UTC, the forecast lead time either coincides with the time of observation or has a shift of 24 hours. Hence, all scores are reported as functions of the lead time. We further average over results from all seven observation locations over Hungary. 

\begin{figure}
  \centering
\epsfig{file=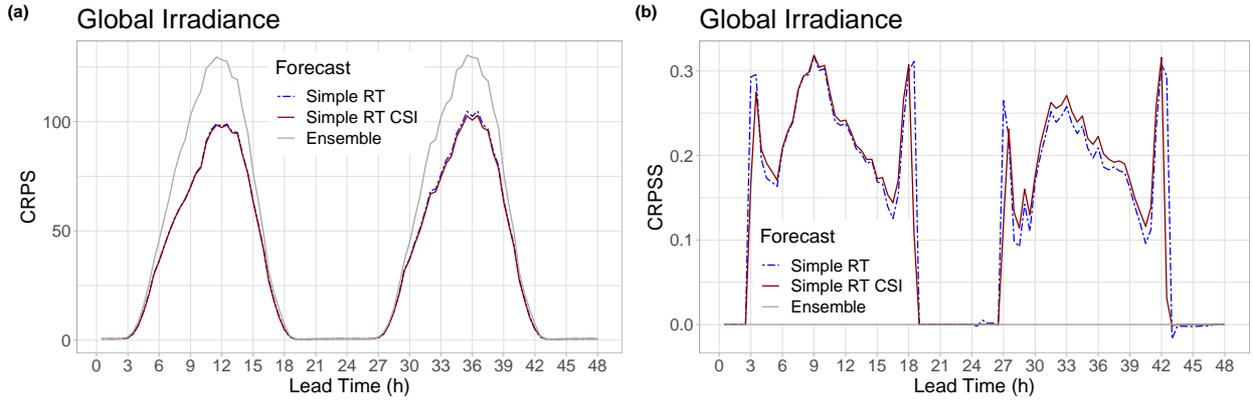, width=\textwidth}
\caption{Mean CRPS of post-processed and raw ensemble forecasts of GHI (a) and CRPSS with respect to the raw ensemble (b) as functions of lead time for the AROME-EPS dataset.}
\label{fig:arome_crps_lead}
\end{figure}

Note that in contrast to considering predictions of GHI directly, the standard approach in solar forecasting is the use of a clear-sky index (CSI) as target variable to stationarize the time series of irradiances \citep{vanderMeerEtAl2018,Yang2020clearsky,YangEtAl2020verification}. The clear-sky irradiance used for the normalization is obtained from clear-sky models which estimate the amount of solar radiation arriving at the surface under clear-sky (cloud-free) conditions, see \citet{Yang2020clearsky} for an in-depth discussion and comparison of available models. To investigate the differences between post-processing forecasts of GHI and forecasts of CSI, we follow the procedure outlined in \citet[][Section II.A]{Yang2020siteadaptation2} to convert the GHI ensemble predictions \ $f_1, f_2, \ldots, f_K$ \ and the GHI observation $\ y\ $ to CSI values. To do so, we obtained clear-sky irradiance values from the McClear model using the API provided by the \texttt{R} package \texttt{camsRad} \citep{camsRad} for the locations and relevant time instances, and converted GHI to CSI by division by the corresponding clear-sky irradiance. We then used an identical model formulation and training procedure as for GHI, and derived 100 equidistant quantiles from the post-processed forecast distributions for CSI. Those quantiles were transformed back to GHI values by multiplying with the corresponding clear-sky irradiance and used for approximating the verification scores. We refer to this approach as {\em simple RT CSI}.

\begin{figure}[p]
  \centering
\epsfig{file=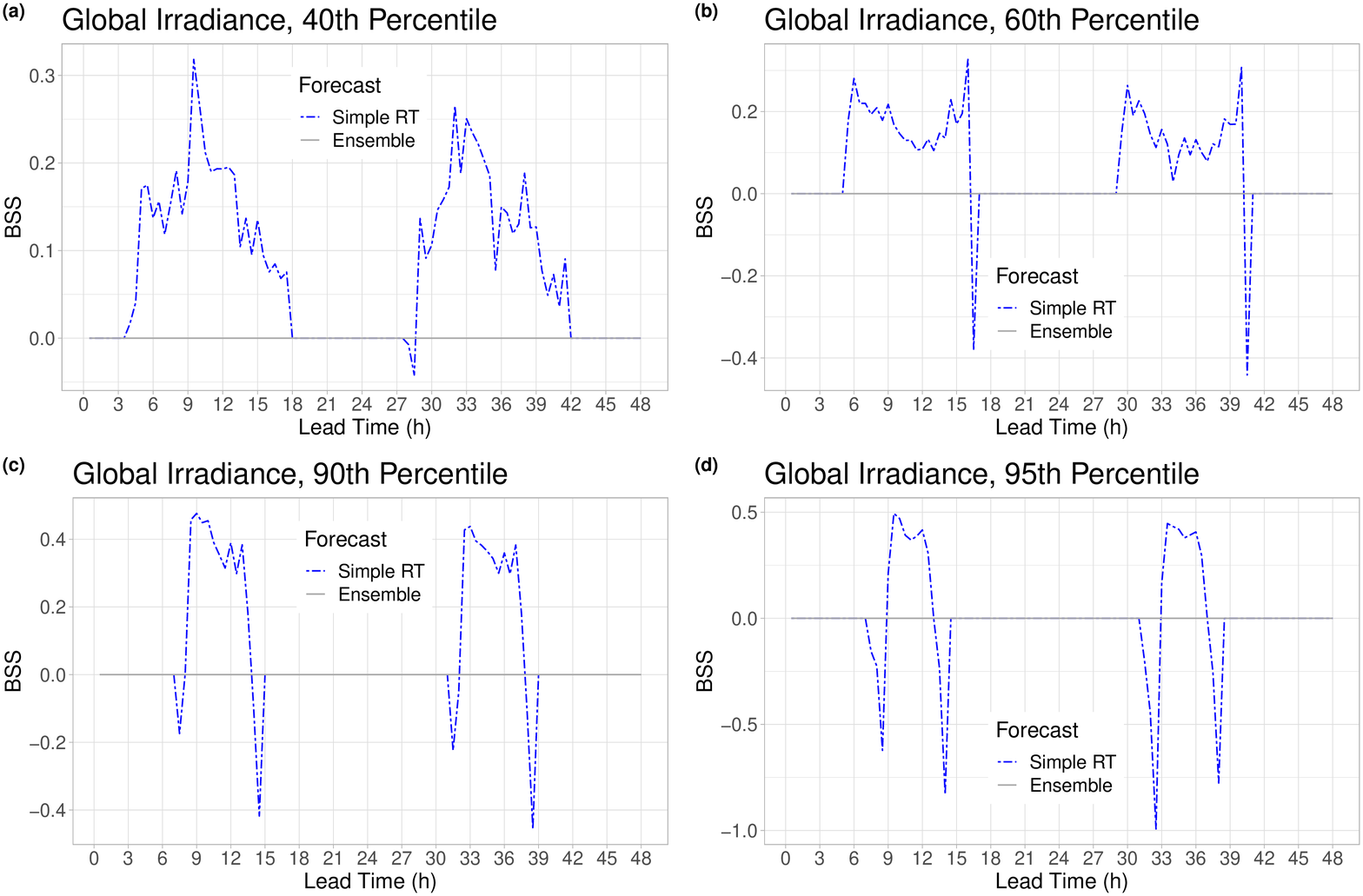, width=\textwidth}
\caption{BSS of post-processed forecasts with respect to the raw ensemble as function of lead time for the AROME-EPS dataset, with thresholds corresponding to the 40th, 60th, 90th and 95th percentiles of observed non-zero GHI.}
\label{fig:arome_bss_lead}

\bigbreak\bigbreak\bigbreak

\epsfig{file=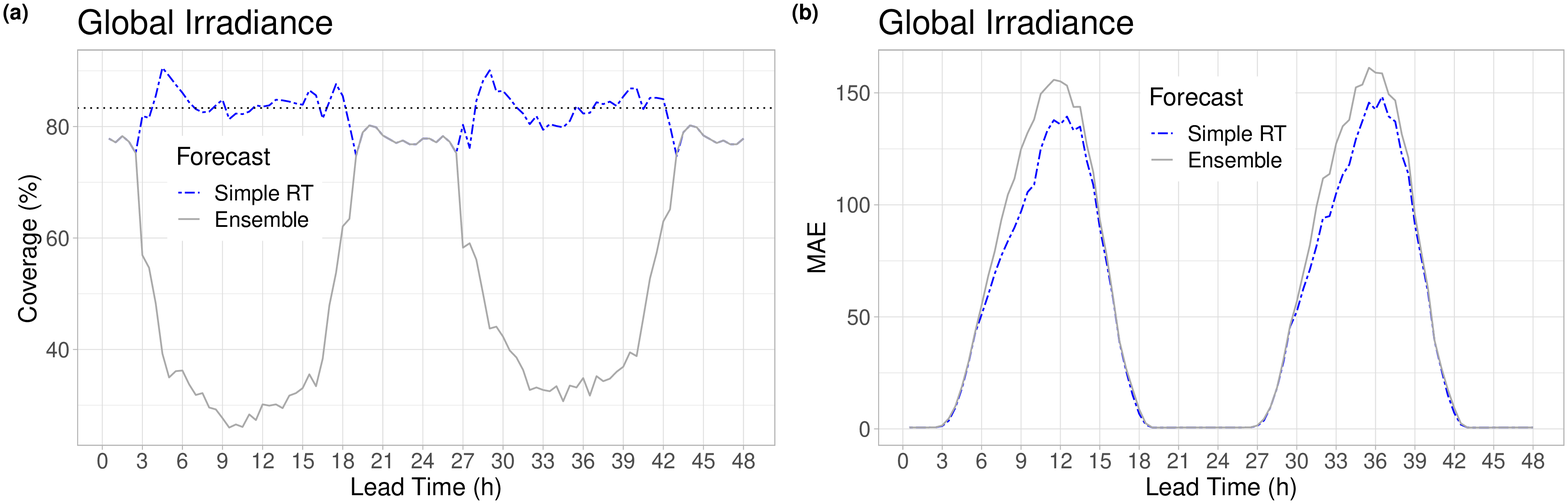, width=\textwidth}
\caption{Coverage of the nominal 83.33\,\% central prediction intervals of post-processed and raw forecasts (a); MAE of the median forecasts (b) for the AROME-EPS dataset.}
\label{fig:arome_cov_lead}
\end{figure}

Figure \ref{fig:arome_crps_lead} shows the mean CRPS of calibrated and raw ensemble forecasts and the CRPSS with respect to the raw ensemble. Post-processing using the simple RT approach improves the forecast performance when positive irradiance is likely to be observed (3 -- 19 UTC), and performs no corrections otherwise, resulting in a skill score of zero. Note that compared with direct calibration of the GHI, post-processing of the CSI predictions does not result in a substantial difference or clear improvement in forecast skill. These observations are in line with the results reported in \citet{Yang2020siteadaptation2} in a related context. Hence, in the remainder only the results for the former approach will be reported. To assess the statistical significance of the improvements in predictive performance compared to the raw ensemble predictions, we performed a block bootstrap resampling to compute 95\% confidence intervals and found that the observed improvements are statistically significant, see Section \ref{subsA.2} of the Appendix. The large jumps in the CRPSS at 4, 19, 27 and 42h are mainly caused by numerical issues as at these lead times the mean CRPS of both raw and post-processed forecasts is very close to $0$, and also leads to an increased width of the confidence intervals computed in Section \ref{subsA.2}. For qualitatively similar observations in a related context, see e.g. \citet[][Figure 7]{BakkerEtAl2019}. An assessment of the improvements in comparison to a climatological reference forecast is provided in Section \ref{secB} of the Appendix.

A similar behavior can be observed for the BSS values shown in Figure \ref{fig:arome_bss_lead}, where the threshold values for $\ z\ $ correspond to the 40th, 60th, 90th and 95th percentiles of observed non-zero GHI (25, 127, 498, 604 $W/m^2$). The results are consistent in that the higher the threshold, the shorter the period with a positive mean BS, as the higher thresholds are mostly observed around midday, when the irradiance is strongest. For the corresponding lead times, the post-processed forecasts outperform the raw ensemble. Again, negative skill scores appear only at the boundaries where the mean score values to be compared are very small.

Figure \ref{fig:arome_cov_lead}a showing the coverage of the nominal 83.33\,\% central prediction intervals further confirms the improved calibration of the post-processed forecast. Between 3 and 19 UTC, when positive GHI is likely to be observed, the EMOS model results in a coverage close to the nominal value, whereas the coverage of the raw ensemble is consistently below 60\%.

Further, Figure \ref{fig:arome_cov_lead}b showing the MAE of the median forecasts indicates that post-processing substantially improves the accuracy of point forecasts as well. At the hours of peak irradiance the difference in MAE exceeds 20 $W/m^2$. As we will see below, this is in a strong contrast with the results of the second case study (see Figure \ref{fig:icon_mae_lead}) and indicates the presence of a bias in the AROME-EPS that is alleviated by post-processing. Similar conclusions can be drawn from the RMSE of the mean forecasts (not shown).

\begin{figure}
  \centering
\epsfig{file=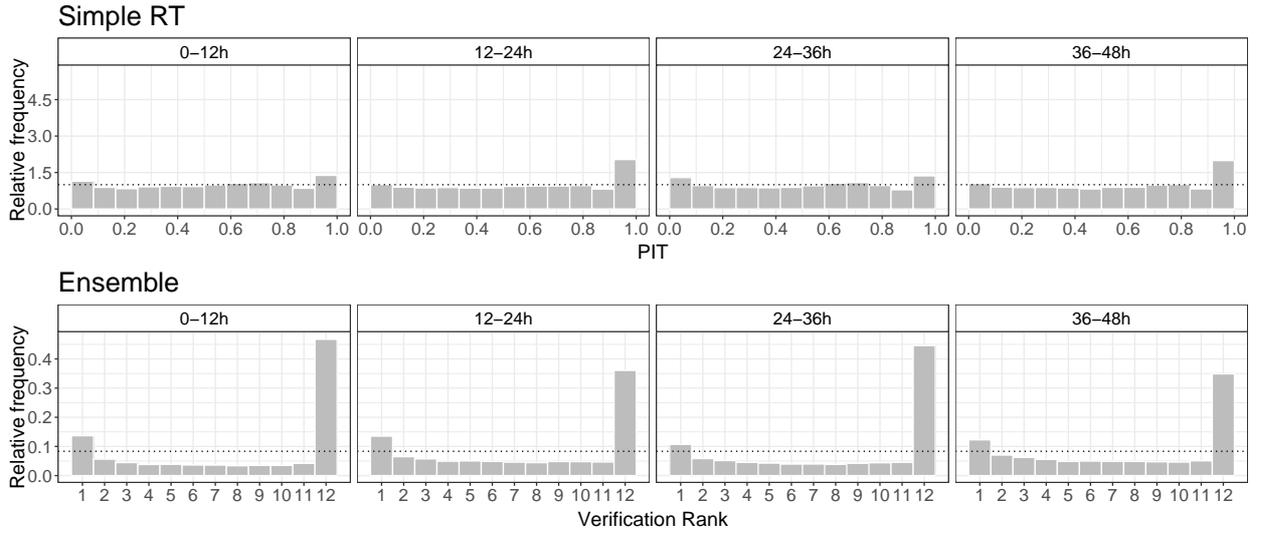, width=\textwidth}
\caption{PIT histograms of post-processed and verification rank histograms of raw ensemble forecasts of GHI for the lead times 0-12h, 12-24h, 24-36h and 36-48h.}
\label{fig:arome_pit}
\end{figure}

The presence of a bias in the raw ensemble forecasts can also be observed in the verification rank histograms shown in Figure \ref{fig:arome_pit}. In addition, the clearly U-shaped verification rank histograms indicate a strong underdispersion, which is in line with the low coverage of the raw ensemble forecasts observed in Figure \ref{fig:arome_cov_lead}a and persists across all considered ranges of lead times. However, the ensemble members are more likely to underestimate the true irradiance, which further indicates a negative bias. Both deficiencies are successfully corrected by statistical post-processing. The PIT histograms of EMOS predictive distributions given in the upper row of  Figure \ref{fig:arome_pit} are almost flat indicating just a minor bias for observations between 12:30 and 24 UTC.

\subsection{Results for the ICON-EPS dataset}
\label{subs4.2}
In contrast to the first case study, the ICON-EPS dataset covers a substantially longer time period and therefore allows for considering and comparing more complex model formulations and estimation procedures. Recall that we here consider forecasts of direct irradiance (BHI) and diffuse irradiance (DHI) at temporal resolutions of 1h (for lead times up to 48h), 3h (for lead times 51 -- 72h) and 6h (for lead times 78 -- 120h), resulting in a forecast horizon of 120h. 

As members of the ICON-EPS are obtained with the help of random perturbations, they can be regarded as exchangeable. Hence, for post-processing we use the \clzero-EMOS model with locations linked to the ensemble members either via \eqref{eq:cl0link1Ex} or via \eqref{eq:cl0link2Ex} with \ $K=1$. \ Thus, for  model \eqref{eq:cl0link1} with location \eqref{eq:cl0link1Ex} (which was the only model variant considered in Section \ref{subs4.1} and is referred to as {\em simple model\/}) one has to estimate five unknown parameters, whereas more complex approaches, which account for seasonal variations in the link function \eqref{eq:cl0link2Ex} of the location parameter via \eqref{eq:reg1} (referred to as {\em periodic model\/}) or \eqref{eq:reg2} (referred to as {\em periodic 2 model\/}), require the estimation of a total of 11 and 15 parameters, respectively. 

The period from 27 December 2018 - 31 December 2019 is used for training purposes only, the calendar year 2020 (366 calendar days) for model verification, which leaves enough flexibility for choosing a sufficiently long training period even for local modeling. Two different training configurations are investigated: a rolling training (RT) period of length 365 days, and a monthly expanding training (MET) scheme, where all data until the end of the last month before the forecast date under consideration is used for training. In the latter case, the first training period includes all data prior to calendar year 2020. According to initial studies (not shown), MET provides reasonable verification scores only for the simple model. Therefore, we report results for the simple model with rolling ({\em simple RT\/}) and monthly expanding training ({\em simple MET\/}) as well as for the periodic models with rolling training ({\em periodic RT} and {\em periodic 2 RT}). Given the negligible differences we observed when comparing post-processing of GHI and CSI for the AROME-EPS data in Section \ref{subs4.1}, we only consider predictions of BHI and DHI without normalization by the corresponding clear-sky irradiances here.

\begin{figure}[t]
  \centering
\epsfig{file=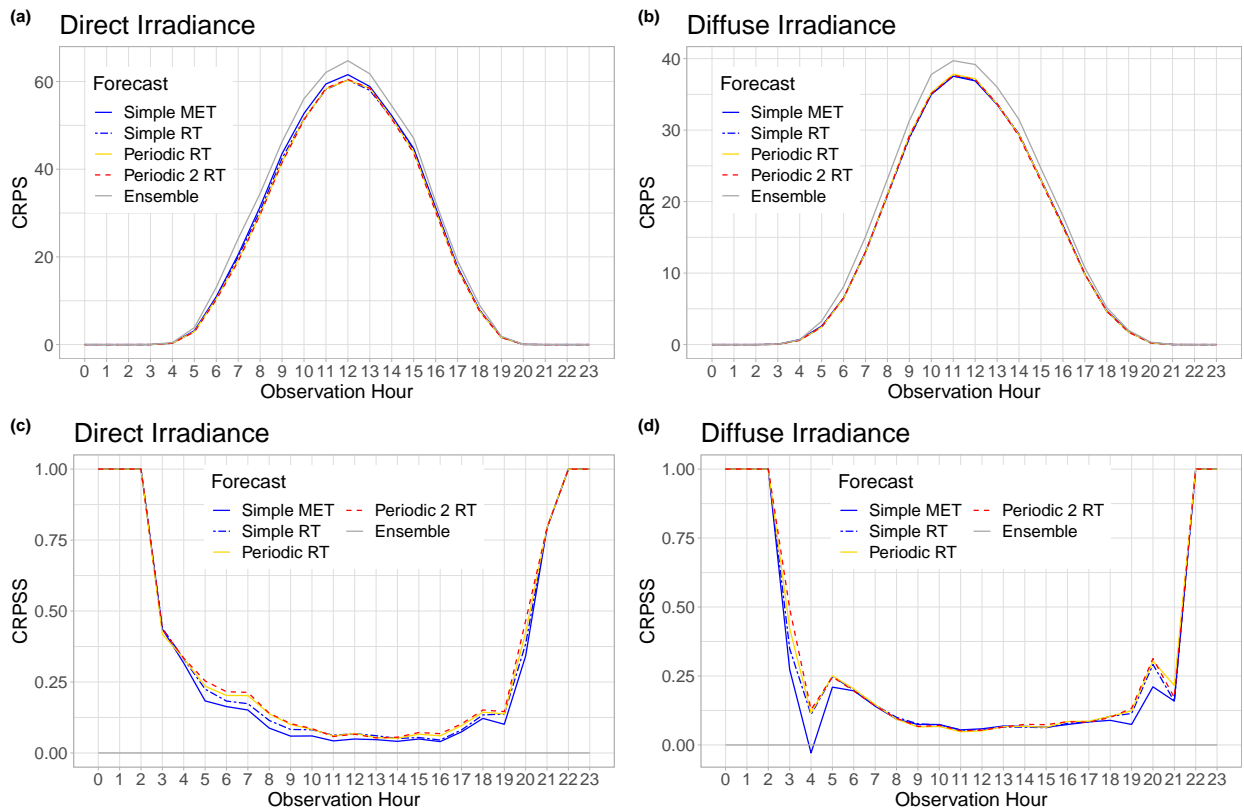, width=\textwidth}
\caption{Mean CRPS of post-processed and raw ensemble forecasts of direct (a) and diffuse (b) irradiance, and corresponding skill scores (c,d) with respect to the raw ensemble as functions of the observation hour for the ICON-EPS dataset.}
\label{fig:icon_crps48}
\end{figure}
%\bigbreak\bigbreak\bigbreak

Raw ensemble forecasts of direct and diffuse irradiance are used as references models. Unless indicated otherwise, results discussed below are averaged over all three observation locations and all four initialization times of the NWP model. Note that this might make the interpretation of the results more involved than in the first case study due to the interacting effects of forecast initialization time, lead time, and corresponding time of day of the observation.

First, we investigate diurnal effects by examining the dependence of the mean CRPS of the various forecast models on the time of the observation shown in Figure \ref{fig:icon_crps48}a,b. 
In order to provide a fair comparison, we take only the first 48h of the forecast horizon into account, where hourly forecasts are available.
Both for BHI and DHI, all post-processing methods outperform the raw ensemble forecasts at all time points when positive irradiance is likely to be observed. According to the skill scores with respect to the raw ensemble shown in Figure \ref{fig:icon_crps48}c, in the case of direct irradiance the predictive performance mainly depends on the complexity of model formulations and parameter estimation, with more complex models exhibiting better forecast performance. However, the differences between the various EMOS approaches are relatively minor. The same applies for diffuse irradiance in early and late hours (see Figure \ref{fig:icon_crps48}d), whereas between 6 -- 18 UTC there is no visible difference in the skill of the different EMOS models. For corresponding results against a climatological reference model, see Section \ref{secB} of the Appendix. 

\begin{figure}[t]
  \centering
\epsfig{file=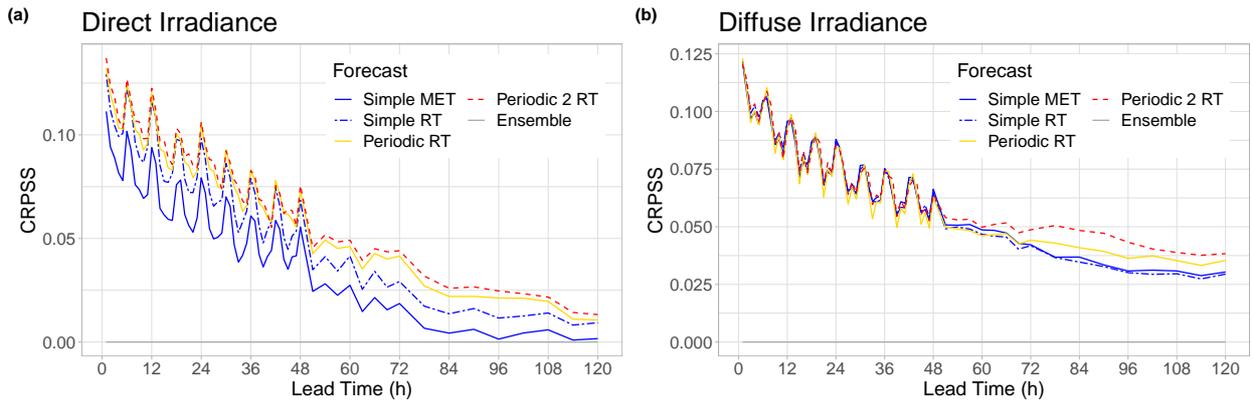, width=\textwidth}
\caption{CRPSS of post-processed forecasts of direct (a) and diffuse (b) irradiance with respect to the raw ensemble as functions of the lead time for the ICON-EPS dataset.}
\label{fig:icon_crps_lead}
\end{figure}

Note that the apparent periodic oscillations in the CRPSS values might be partly caused by the pooling of different observation hours due to the four considered initialization times. In contrast to the AROME-EPS, post-processing also improves the predictive performance at night achieving a CRPS of almost zero. ICON-EPS fails to achieve mean CRPS values of zero due to occasional predictions of non-zero irradiance values during night times.

\begin{table}[t]{\scriptsize
    \begin{center}
  \begin{tabular}{l|l|c|c|c|c|c|c|c|c}
    Lead&\multicolumn{1}{c|}{Model}&\multicolumn{4}{c|}{Direct Irradiance}&\multicolumn{4}{c}{Diffuse Irradiance} \\ \cline{3-10} Time&&Overall&Karlsruhe&Berlin&Hamburg&Overall&Karlsruhe&Berlin&Hamburg \\ \hline
        &Simple MET&0.076&0.114&0.029&0.082&0.089&0.099&0.075&0.093\\
    1-24h&Simple RT&0.095&0.130&0.070&0.083&0.091&0.100&0.080&0.092\\   &Periodic RT&0.101&0.133&0.077&0.090&0.089&0.099&0.080&0.086\\
        &Periodic 2 RT&0.104&0.132&0.082&0.097&0.091&0.097&0.083&0.092\\ \hline
        &Simple MET&0.050&0.091&-0.000&0.054&0.066&0.076&0.050&0.070\\
  25-48h&Simple RT&0.064&0.102&0.032&0.054&0.066&0.075&0.053&0.069\\
        &Periodic RT&0.072&0.102&0.045&0.066&0.064&0.071&0.055&0.063\\
        &Periodic 2 RT&0.074&0.099&0.047&0.073&0.066&0.069&0.060&0.069\\ \hline
        &Simple MET&0.021&0.062&-0.018&0.018&0.048&0.055&0.041&0.046\\
  51-72h&Simple RT&0.033&0.071&0.008&0.020&0.046&0.053&0.041&0.043\\
        &Periodic RT&0.043&0.075&0.021&0.030&0.047&0.053&0.045&0.041\\
        &Periodic 2 RT&0.046&0.069&0.024&0.043&0.051&0.049&0.055&0.049\\ \hline
       &Simple MET&0.004&0.028&-0.016&-0.002&0.032&0.032&0.039&0.025\\
 78-120h&Simple RT&0.013&0.036&-0.001&0.001&0.031&0.031&0.040&0.021\\
        &Periodic RT&0.019&0.032&0.017&0.008&0.038&0.037&0.048&0.026\\
        &Periodic 2 RT&0.022&0.024&0.021&0.022&0.043&0.036&0.062&0.030\\ \hline
  \end{tabular}
\caption{Overall CRPSS and CRPSS for individual locations of post-processed forecasts of direct and diffuse irradiance with respect to the raw ensemble.}
\label{tab:crpss}
\end{center}}
\end{table}

Figure \ref{fig:icon_crps_lead}a, which shows the CRPSS with respect to the raw BHI ensemble forecasts as function of the lead time, confirms the observations from Figure \ref{fig:icon_crps48}c. Here the differences are more pronounced due to the different scaling of the vertical axis, and again, the periodic 2 model with rolling training period exhibits the best forecast skill, whereas the simple model with monthly expanding training shows the smallest CRPSS. In general, all skill scores decrease for longer lead times, which also holds for the corresponding CRPSS values for DHI (Figure \ref{fig:icon_crps_lead}b). Overall, slightly larger improvements relative to the raw ensemble are observed for direct than for diffuse irradiance, and none of the models result in negative skill scores. Up to a lead time of 48h, there are no visible differences between the various EMOS approaches. For longer lead times, similar to BHI, the most complex periodic 2 model shows the best predictive performance, whereas the simple model with parameters estimated using a rolling training period is now the least skillful. Recall that for longer lead times the forecasts refer to a longer time period and thus seasonal effects regarding the diurnal cycle might be captured by the more complex models.

An assessment of the statistical significance of the improvements in predictive performance compared to the raw ensemble predictions shows that in the case of BHI, post-processing results in a significant improvement in mean CRPS up to 60h ahead, whereas post-processed forecasts of DHI significantly outperform the raw ensemble over the entire forecast horizon of 120h. The corresponding results and more details can be found in Section \ref{subsA.1} of the Appendix.

\begin{figure}[t]
  \centering
\epsfig{file=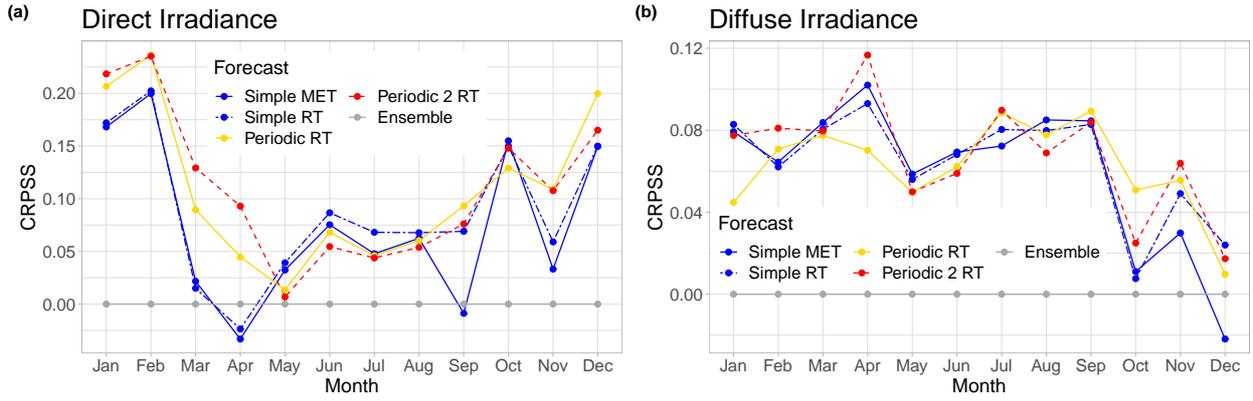, width=\textwidth}
\caption{CRPSS of post-processed forecasts of direct (a) and diffuse (b) irradiance with respect to the raw ensemble, computed based on monthly mean values for the ICON-EPS dataset.}
\label{fig:icon_crpss_month}
\end{figure}

A third aspect is the dependence of the forecast skill on the observation location. Table \ref{tab:crpss} shows the overall CRPSS of the different EMOS models with respect to the raw forecasts and the corresponding CRPSS values of the three different cities for four different intervals of the forecast horizon. The main message of these results is that the magnitude of improvements in predictive performance resulting from post-processing strongly depends on the location. For both variables, Karlsruhe benefits the most, while for Berlin after 24h, and for Hamburg after 78h the simple MET model performs worse than the raw BHI ensemble forecast and results in negative skill scores. Among the competing models for BHI the most complex periodic 2 RT model shows the best forecast skill for Berlin and Hamburg, and shows the best overall performance as well. In the case of DHI, the differences in performance between the various EMOS models are much smaller, which is in line with the results observed in Figure \ref{fig:icon_crps_lead}b. In particular, none of the more complex models consistently outperforms the simple MET model.

To investigate seasonal effects in the improvements achieved via post-processing, Figure \ref{fig:icon_crpss_month} shows the CRPSS of the post-processed forecasts based on monthly mean values. For direct irradiance, the improvements are generally larger in winter than in summer. From November to April, the differences among the post-processing approaches are most pronounced, and more complex model formulations that incorporate seasonal effects particularly show improved performance. For diffuse irradiance, the overall level of improvements in terms of the mean CRPS is smaller (note the different scale of the vertical axes). Only minor seasonal effects in the form of smaller improvements between October and December can be detected.

\begin{figure}[t]
  \centering
\epsfig{file=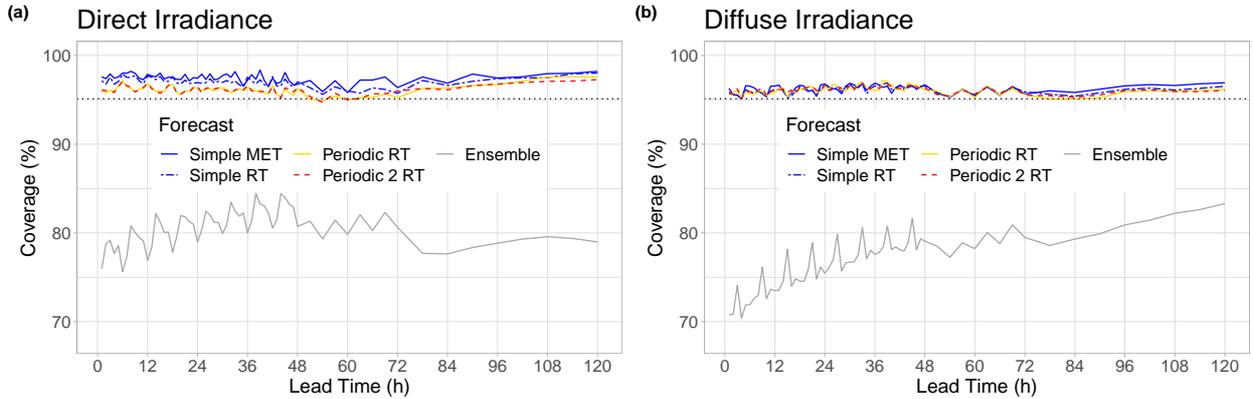, width=\textwidth}
\caption{Coverage of nominal $95.12\,\%$ central prediction intervals of post-processed and raw forecasts of direct (a) and diffuse (b) irradiance for the ICON-EPS dataset.}
\label{fig:icon_cov_lead}
\end{figure}

To simplify the presentation of the results, in the remaining part of this section we consider pooled data of all locations, months and observation hours and display the dependence on the lead time only.

\begin{figure}
  \centering
\epsfig{file=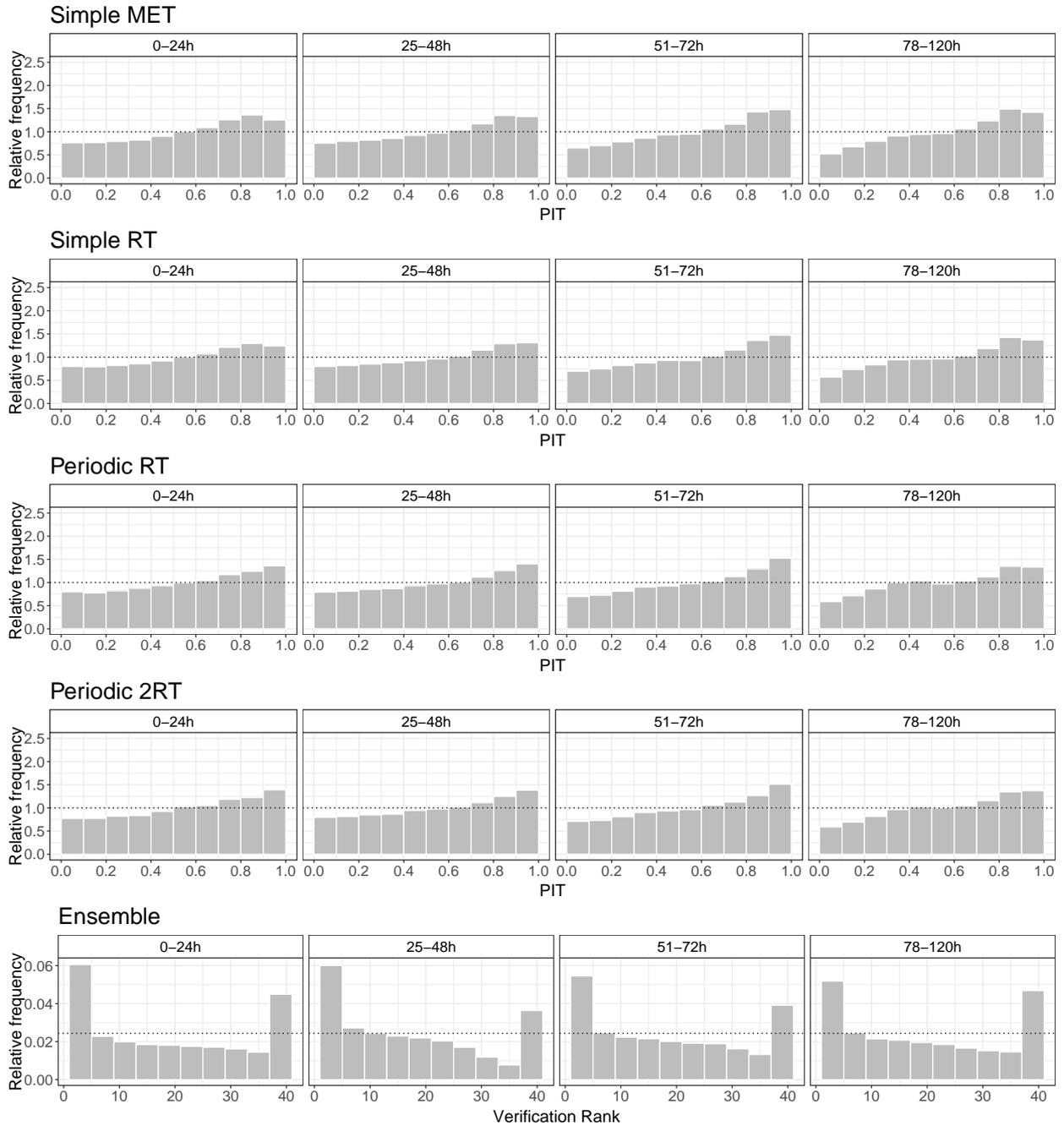, width=\textwidth}
\caption{PIT histograms of post-processed and verification rank histograms of raw ensemble forecasts of DNI for the lead times 0-24h, 25-48h, 51-72h and 78-120h for the ICON-EPS dataset.}
\label{fig:icon_pit_dir}
\end{figure}

\begin{figure}
  \centering
\epsfig{file=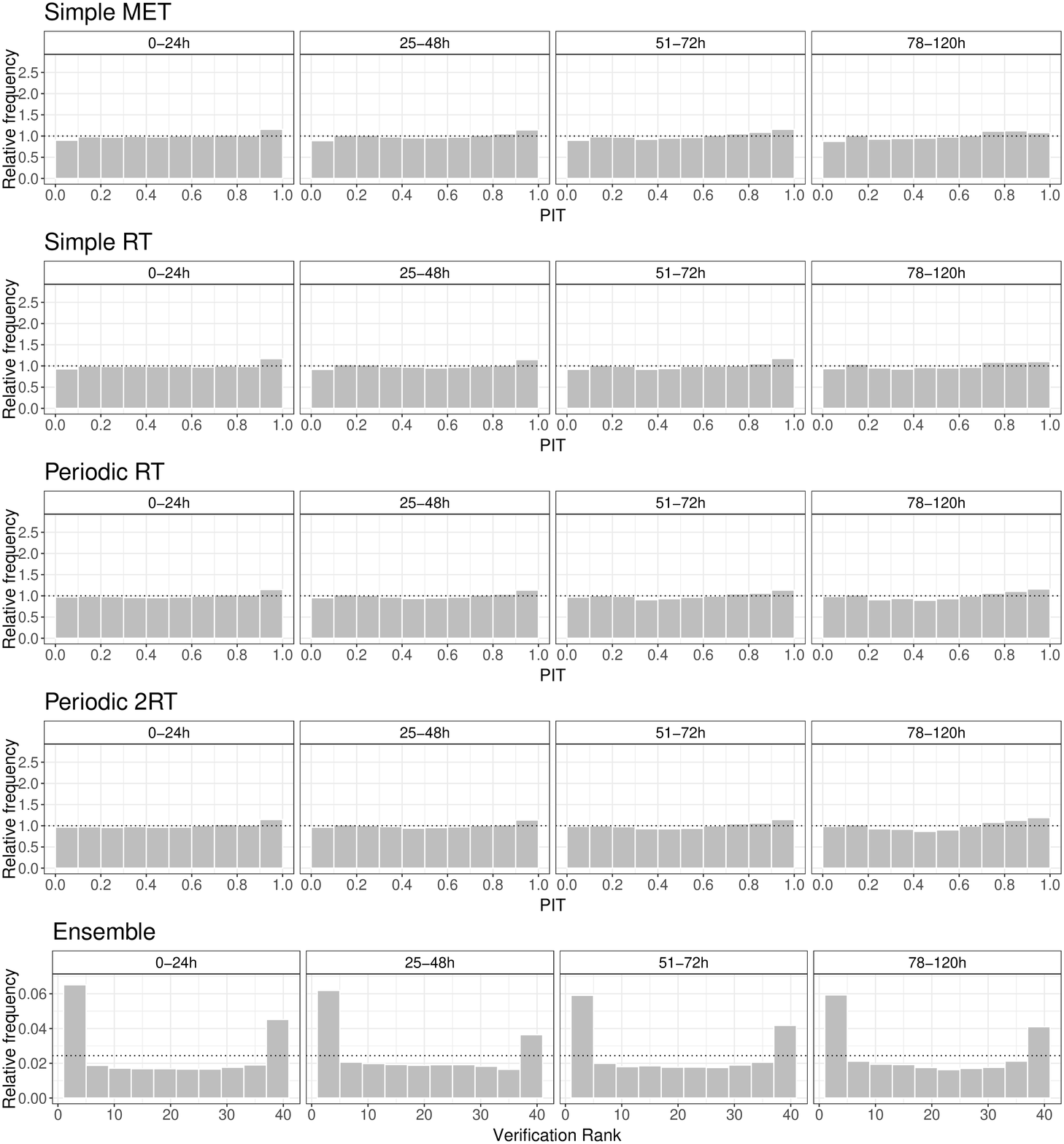, width=\textwidth}
\caption{PIT histograms of post-processed and verification rank histograms of raw ensemble forecasts of DHI for the lead times 0-24h, 25-48h, 51-72h and 78-120h for the ICON-EPS dataset.}
\label{fig:icon_pit_difd}
\end{figure}

The improved calibration of post-processed forecasts can also be observed in the coverage plots of Figure \ref{fig:icon_cov_lead}. All post-processing approaches result in a coverage close to the nominal $95.12\,\%$ for all lead times, whereas the maximal coverage of the raw ensemble is below $85\,\%$ for both variables. The difference between post-processed forecasts of BHI is more pronounced with periodic models being the closest to the nominal value. These results are in line with the shapes of the PIT and verification rank histograms of Figures \ref{fig:icon_pit_dir} and \ref{fig:icon_pit_difd}. Raw ensemble forecasts of BHI are strongly underdispersive and slightly biased for all lead times. Even though this is slightly alleviated for longer lead times, the PIT histograms of all EMOS models are much closer to the desired uniform distribution. However, some bias still remains in the post-processed forecasts. In contrast, neither the PIT histograms of post-processed, nor the verification rank histograms of raw forecasts of DHI indicate any bias (Figure \ref{fig:icon_pit_difd}), and all EMOS approaches successfully correct the underdispersion of the raw ensemble resulting in almost perfectly uniform PIT histograms.

\begin{figure}
  \centering
\epsfig{file=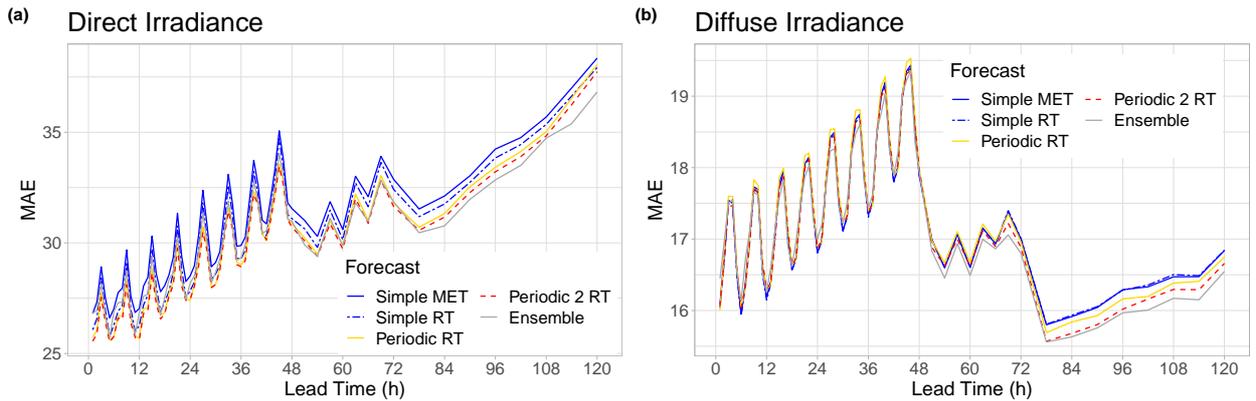, width=\textwidth}
\caption{MAE of the median forecasts of direct (a) and diffuse (b) irradiance.}
\label{fig:icon_mae_lead}
\end{figure}

Finally, Figure \ref{fig:icon_mae_lead} showing the MAE of the median forecasts indicates that while post-processing substantially improves the calibration of probabilistic forecasts, it has a minor effect on the accuracy of point forecasts. The difference in MAE is less than 2 $W/m^2$ for direct irradiance and 0.6 $W/m^2$ for diffuse irradiance for all considered lead times. The sharp changes in MAE values at 51h and 78h are results of the change in temporal resolution of the forecasts. Corresponding results for the RMSE of the mean forecasts are very similar and thus not shown here.

\section{Conclusions}
\label{sec5}

We propose a post-processing method for ensemble weather predictions of solar irradiance where probabilistic forecasts are obtained in the form of a logistic distribution left-censored at zero. Several model variants that differ in terms of the temporal composition of training datasets and adjustments to seasonal variations in the model formulation are evaluated in two case studies. Even though the case studies cover distinct geographical regions, NWP systems, types of solar irradiance and temporal resolutions, the results presented in Section \ref{sec4} indicate that the proposed post-processing models are able to consistently and significantly improve the forecast performance of the raw ensemble predictions up to lead times of at least 48 hours. The improvements from post-processing are larger for the AROME-EPS dataset, possibly due to a lower skill of the raw ensemble predictions resulting from a bias in addition to the observed underdispersion. For the ICON-EPS dataset, we observed that more complex post-processing models tend to show better predictive performances, but the differences between model variations rarely show a high level of statistical significance. For the GHI predictions of the AROME-EPS dataset, we only found negligible differences when comparing post-processing models for GHI and CSI. This is in line with the results reported in \citet{Yang2020siteadaptation2} and suggests that the standard practice of normalizing the irradiance forecasts by clear-sky irradiance does not lead to improvements in forecast performance here.

The overall level of improvements achieved via statistical post-processing of the solar irradiance forecasts of the raw ensemble are comparable to meteorological variables such as precipitation accumulation \citep{sch14,bn16} or total cloud cover \citep{BaranEtAl2020} in case of the ICON-EPS dataset, and slightly larger for the AROME-EPS data. Post-processing ensemble predictions of those variables is often seen as a more difficult task compared to variables such as temperature \citep{GneitingEtAl2005} or wind speed \citep{tg10} where substantially larger improvements can be achieved.
Nonetheless, the observed improvements are statistically significant for lead times of up to 2 days, and will likely be of relevance for solar energy forecasting in terms of potential economic benefits and improved balancing of demand and supply for integrating volatile PV power systems into the electrical grid. The datasets used in the two case studies are somewhat limited in terms of their temporal extent, in particular the AROME-EPS data. An interesting aspect for future work will be to for example compare different ways of accounting for seasonal variability once longer, ideally multi-year periods of data have become available.

The post-processing models for solar irradiance considered here provide several avenues for future work. Regarding the general model setup, we have only used ensemble predictions of the target variable as inputs to the post-processing model. However, a recent focus of the post-processing literature has been the use of modern machine learning methods that allow one to incorporate arbitrary predictor variables and to model possibly nonlinear relations to the forecast distribution parameters \citep[see][for a recent review]{VannitsemEtAl2020}. The EMOS models considered here could for example be extended following the gradient boosting extension framework proposed in \citet{MessnerEtAl2017} or the neural network-based approach of \citet{RaspLerch2018}. For similar considerations in the solar irradiance forecasting literature where additional predictors from NWP model output are used, albeit for different types for probabilistic forecasting methods, see, for example, \citet{SperatiEtAl2016} and \citet{BakkerEtAl2019}. 

Further, we have restricted our attention to univariate forecasts for a single location, lead time and target variable. However, in particular in the context of energy forecasting, many practical applications require an accurate modeling of spatial, temporal, or inter-variable dependencies \citep{PinsonMessner2018}. A large variety of multivariate post-processing methods has been proposed over the past years \citep[see][for a recent overview]{LerchEtAl2020}, and a comparison of those approaches in the context of solar energy forecasting might be an interesting starting point for future research. 

Finally, the development of post-processing models for solar irradiance was motivated by the aim of improving probabilistic solar energy forecasting. To that end, it would be interesting to investigate the effect of post-processing NWP ensemble forecasts of solar irradiance for PV power prediction, and for example compare to direct probabilistic models of PV power output \citep[see e.g.][]{AlessandriniEtAl2015}. In a related study in the context of wind energy, \citet{PhippsEtAl2020} find that a two-step strategy of post-processing both wind and power ensemble forecasts performs best and that the calibration of the power predictions constitutes a crucial step. Ideally, statistical post-processing of solar irradiance forecasts could contribute an important component to modern, fully integrated renewable energy forecasting systems \citep[see e.g.][]{HauptEtAl2020}.

\bigskip
\noindent
{\bf Acknowledgments.} \ Benedikt Schulz and Sebastian Lerch gratefully acknowledge support by the Deutsche Forschungsgemeinschaft through SFB/TRR 165 ``Waves to Weather''.  S\'andor Baran was supported by the National Research, Development and Innovation Office under Grant No.\ NN125679. He also acknowledges the support of the EFOP-3.6.2-16-2017-00015 project, which was co-financed by the Hungarian Government and the European Social Fund. 
The authors thank Gabriella Sz\'epsz\'o and Mih\'aly Sz\H ucs from the HMS for providing the AROME-EPS data. We thank DWD for providing access to ICON-EPS and station observation data via the Open Data Server of DWD (\url{https://opendata.dwd.de/}), and Marco Wurth for assistance in obtaining the forecast data. Constructive comments on an earlier version of the manuscript by two anonymous referees are gratefully acknowledged.

\bibliographystyle{myims2}
\bibliography{RadiationPaperArXiv2}

\appendix
\section*{Appendix}

\section{Significance of improvement in forecast performance}
\label{secA}
To get an insight into the uncertainty and statistical significance of the verification scores, we accompany the CRPSS values of the best performing EMOS models (simple RT for the AROME-EPS dataset and periodic 2 RT for the ICON-EPS dataset) with $95\,\%$ confidence intervals. The corresponding standard deviations are obtained from 2\,000 block bootstrap samples calculated using the stationary bootstrap scheme, where the mean block length is computed according to \citet{pr94}.

\begin{figure}[t]
  \centering
\epsfig{file=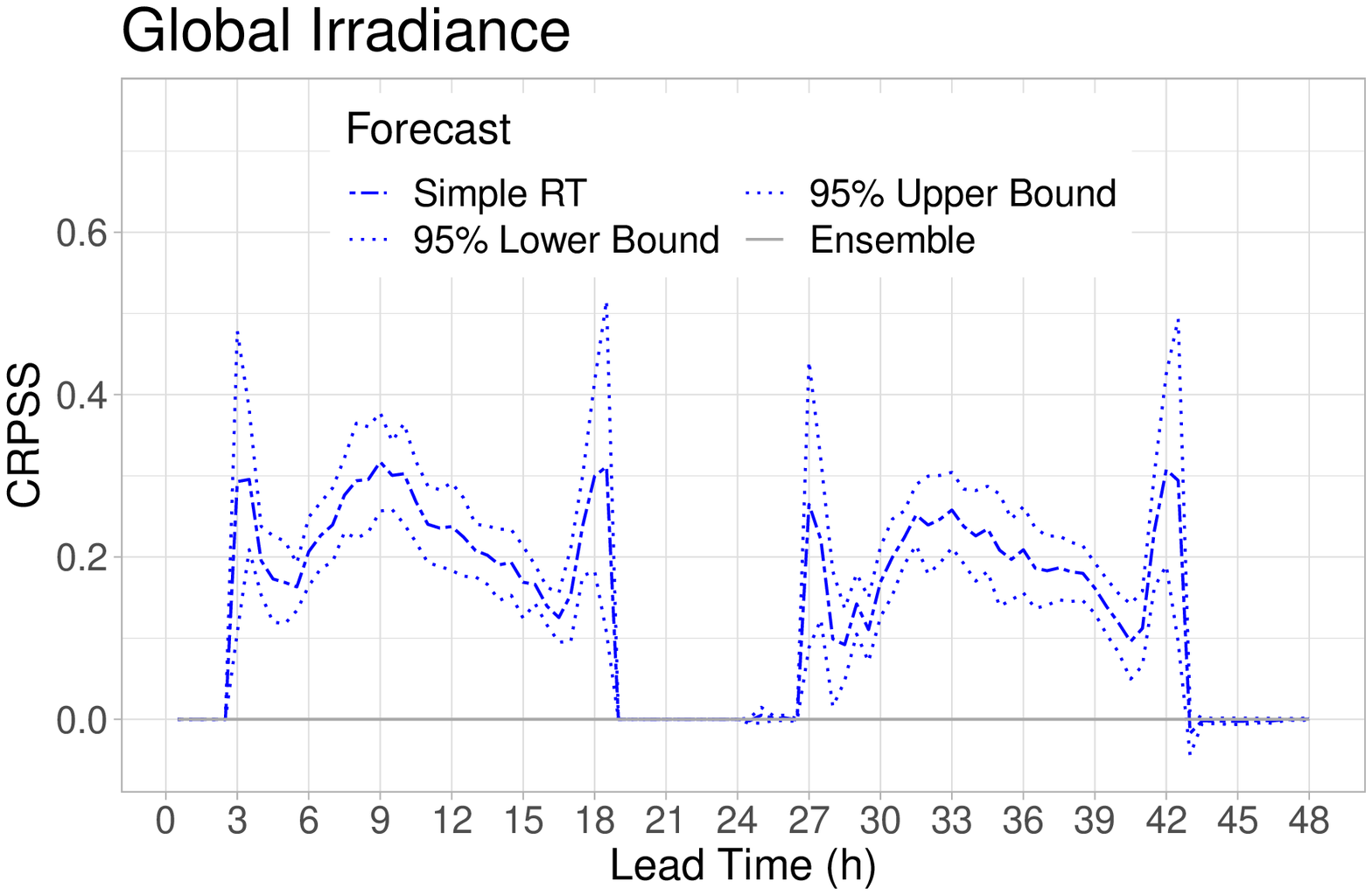, width=0.5\textwidth}
\caption{CRPSS of EMOS post-processed forecasts with respect to the raw ensemble together with $95\,\%$ confidence intervals for the AROME-EPS dataset.}
\label{fig:arome_boot_crps_lead}
%\end{figure}
\bigbreak\bigbreak\bigbreak

%\begin{figure}[t]
  \centering
\epsfig{file=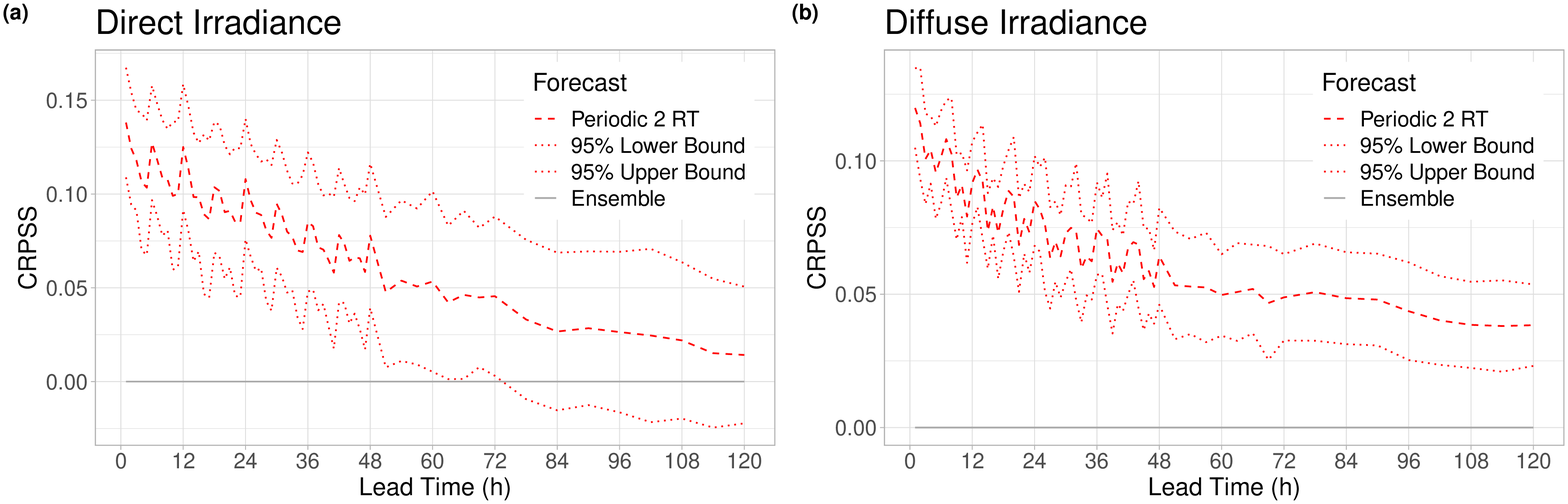, width=\textwidth}
\caption{CRPSS of the best performing post-processed forecasts of direct (a) and (b) diffuse irradiance with respect to the raw ensemble together with $95\,\%$ confidence intervals for the ICON-EPS dataset.}
\label{fig:icon_boot_crps_lead}
\end{figure}

\subsection{CRPSS of the AROME-EPS}
\label{subsA.2}
In the case of the AROME-EPS dataset we consider only the simple EMOS model with a 31 day rolling training period. Figure \ref{fig:arome_boot_crps_lead} is an enhanced version of Figure \ref{fig:arome_crps_lead}b where corresponding $95\,\%$ confidence intervals for the CRPSS are added. The skill scores of post-processed forecasts are significantly positive for all time periods where positive irradiance is likely to be observed.

\subsection{CRPSS of the ICON-EPS}
  \label{subsA.1}

Figure \ref{fig:icon_boot_crps_lead} shows the CRPSS of the best performing periodic 2 RT EMOS model with respect to the raw ICON-EPS forecast and the corresponding $95\,\%$ confidence intervals. In the case of direct irradiance the improvement in mean CRPS is significant up to 60h, whereas for diffuse irradiance it is significant for all considered lead times. Further, comparing Figures \ref{fig:icon_boot_crps_lead} and \ref{fig:icon_crps_lead} it can be observed (especially in the case of diffuse irradiance), that in terms of mean CRPS there is no significant difference between the various post-processing methods.

%\newpage 

\section{Predictive performance with respect to climatology}
\label{secB}

\begin{figure}[t]
  \centering
\epsfig{file=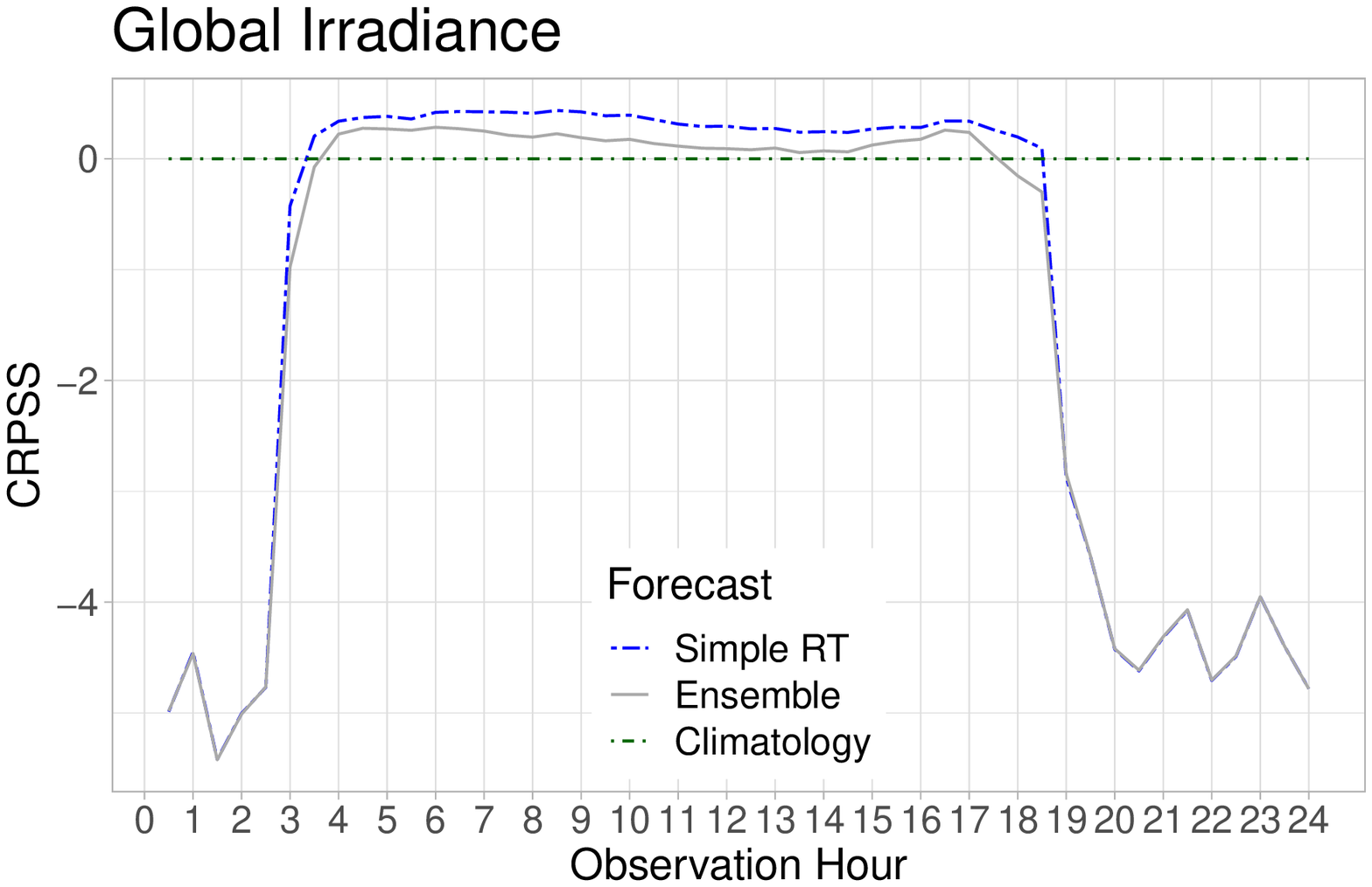, width=0.5\textwidth}
\caption{CRPSS of EMOS post-processed and raw ensemble forecasts with respect to climatology as functions of the observation hour for the AROME-EPS dataset.}
\label{fig:arome_crps_clim}

\bigbreak\bigbreak\bigbreak

\epsfig{file=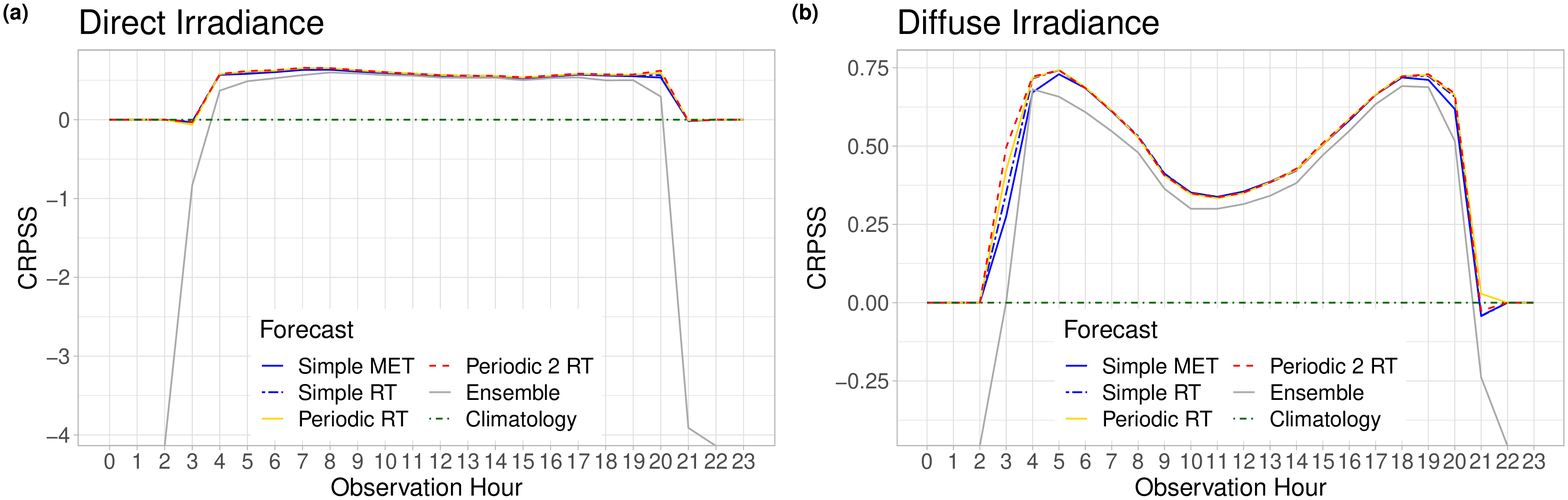, width=\textwidth}
\caption{CRPSS of post-processed and raw ensemble forecasts of direct (a) and (b) diffuse irradiance with respect to climatology as functions of the observation hour for the ICON-EPS dataset.}
\label{fig:icon_crps_clim}
\end{figure}

The relative improvements of the post-processed forecasts in terms of the CRPSS were investigated in Section \ref{sec4} by computing the corresponding skill score values of the CRPSS using the ensemble forecast as reference (see Figures \ref{fig:arome_crps_lead} and \ref{fig:icon_crps48}c,d). To further investigate the relative improvements in comparison to a more na\"{i}ve reference, Figures \ref{fig:arome_crps_clim} and \ref{fig:icon_crps_clim} show corresponding values of the CRPSS using climatological forecast as reference model, where observations of the rolling training period are considered as a forecast ensemble. This can be viewed as a persistence ensemble in the terminology of \citet{Yang2019ROPES}.

In the case of the AROME-EPS individual climatological forecasts are based on observations of the preceding 31 days, whereas for the ICON-EPS a climatological ensemble has 365 members. Both ensemble and the post-processed forecasts show clear improvements for times of day during which it is unlikely to observe zero irradiance. Corresponding skill scores for MAE and RMSE indicate a very similar behavior of the deterministic forecasts and are not shown here in the interest of brevity. 
\end{document}